\documentclass[aps,11pt,preprint]{revtex4}

\usepackage[dvips]{graphicx}
\usepackage{amsfonts}
\usepackage{amssymb,amsmath}
\usepackage{mathrsfs}
\usepackage{bm}
\usepackage{verbatim}

\newcommand{\beq}{\begin{equation}}
\newcommand{\eeq}{\end{equation}}
\newcommand{\bea}{\begin{eqnarray}}
\newcommand{\eea}{\end{eqnarray}}

\newcommand{\reffig}[1]{FIG.~\ref{#1}}

\newcommand{\simlt}{\stackrel{<}{{}_\sim}}

\newcommand{\iv}[1]{\bm{#1}}

\newcommand{\mqcd}{M_{\textrm{QCD}}}

\graphicspath{{../}{figs/}{minuit/plots/}}

\begin{document}

\title{The Role of the Roper in Chiral Perturbation Theory}

\author{Bingwei Long}
\affiliation{Excited Baryon Analysis Center (EBAC), Jefferson Laboratory, 
  12000 Jefferson Avenue, Newport News, Virginia 23606, USA}
\affiliation{European Centre for Theoretical Studies in Nuclear
  Physics and Related Areas (ECT*), I-38123 Villazzano (TN), Italy}

\author{U. van Kolck}
\affiliation{Department of Physics, University of Arizona, Tucson,
  Arizona 85721, USA}
\affiliation{Instituto de F\'{i}sica Te\'orica, Universidade Estadual Paulista,
Rua Dr. Bento Teobaldo Ferraz, 271 - Bloco II,
01140-070, S\~ao Paulo, Brazil}

\date{\today}

\begin{abstract}
We include the Roper excitation of the nucleon in a version of
heavy-baryon chiral perturbation theory recently developed for energies around
the delta resonance.
We find significant improvement in the $P_{11}$ channel.
\end{abstract}

\preprint{JLAB-THY-11-1372}
\preprint{INT-PUB-11-020}


\maketitle

Chiral perturbation theory (ChPT) is the effective field theory (EFT) of QCD 
at momenta comparable to the pion mass, $m_\pi$. 
It includes the lightest hadron, the pion, but not mesons ---such as the rho---
with masses of the order of the typical QCD mass scale, $\mqcd\sim 1$ GeV. 
The mesonic version of the theory \cite{weinberg79} can be thought of as 
an expansion of amplitudes in $m_\pi/\mqcd$ and $Q/\mqcd$, 
where $Q$ is a characteristic external momentum.
This approach has been shown to be very successful 
\cite{BerKaiMei} for a variety of processes at energies below 
a structure associated with the sigma meson at a position
$m_\sigma -i\Gamma_\sigma/2 =(441-272i)$ MeV \cite{Caprini:2005zr}
in the complex energy plane,
which suggests a radius of convergence 
$\Sigma \sim \sqrt{m_\sigma^2 + \Gamma_\sigma^2/4}\simeq 6f_\pi$,
where $f_\pi\simeq 92$ MeV is the pion decay.

ChPT includes also the lightest baryon, the nucleon, because the relatively 
large nucleon mass, $m_N$, is inert in low-energy process \cite{gasser}. 
However, contrary to the mesonic sector, the first baryon excitations appear 
at energies not much larger (relative to $m_N$) than $m_\pi$.
The most important is the delta isobar at 
$m_{\Delta} - m_N -i \Gamma_\Delta/2\simeq (270- 50i)$ MeV \cite{gwpwa}. 
If the delta is not included explicitly, the theory represents an expansion 
in $\sim (Q,m_\pi)/\delta$ with 
$\delta \equiv m_\Delta - m_N \simeq 3 f_\pi$,
which cannot be applied at energies much beyond the threshold region. 
The importance of the delta in ChPT has been recognized for a long time, and
introducing a field to describe its long-distance effects rearranges 
ChPT contributions and improves its convergence pattern  
\cite{jenkins, hemmertdelta, bira-thesis}. 
In order to calculate amplitudes in the vicinity of $\delta$, 
a selective resummation is required \cite{pascalutsa, us}. 
Generally this yields very good results \cite{morepascalutsa1,morepascalutsa2}.
In the quintessential low-energy nucleon reaction, elastic $\pi N$ scattering, 
the leading delta contribution is of $\mathcal{O}(\mqcd^2/Q)$,
in contrast with ChPT near threshold \cite{BerKaiMei} where ``leading order''
is used to refer to $\mathcal{O}(Q)$. 
All channels 
except $P_{11}$ are well described at $\mathcal{O}(Q)$ \cite{us},
which here means next-to-next-to-leading order.

Other nucleon excitations have received considerably less attention in ChPT. 
Among them, the Roper \cite{origroper}
is special, and here we argue that the Roper can be 
considered within the regime of ChPT, although of course in a marginal sense. 
First, the Roper pole appears at an energy not very far above the delta,
$m_{R} - m_N -i\Gamma_R/2\simeq (420 - 80i)$ MeV \cite{gwpwa}.
Without explicit Roper contributions, the theory is as an expansion in 
$\sim (Q, m_\pi)/\rho$, where 
$Q$ now includes $\delta$, and 
$\rho \equiv m_R - m_N \simeq 4.5 f_\pi$.
Other resonances lay at least $\Sigma$
above threshold (for example, the next higher $S_{11}$ resonance  
has a larger mass, $m_{S_{11}} - m_N \gtrsim 6f_\pi \simeq \Sigma$ 
\cite{gwpwa}),
and it is difficult to
see how they could be incorporated in the EFT without
the concomitant inclusion of meson resonances.
Second, the Roper width is $\Gamma_R\sim \Gamma_\Delta \rho^3/2\delta^3$,
as expected from ChPT widths that scale as $Q^3/\mqcd^2$. 
The same is not true for higher resonances,
which typically have smaller relative widths.
Third, the delta and the Roper nearly saturate the Adler-Weisberger 
sum rule, a result which suggests that, together with the nucleon, 
these two resonances fall into a simple reducible representation of the 
chiral $SU(2)_L \times SU(2)_R$ group \cite{algreal,roperQCD}. 

The Roper could thus be expected to play a role in low-energy observables. 
Take, for example, elastic $\pi N$ scattering in the $P_{11}$ channel. 
The phase shift \cite{gwpwa} is repulsive near threshold but becomes 
attractive at a center-of-mass (CM) energy 
(with $m_N$ subtracted out) $E \sim 3 f_\pi$,
right in the delta region.
In ChPT, whether without 
\cite{mojzis, fettes-deltaless-is, otherdeltaless, moreIRdeltaless} 
or with \cite{threshold-delta1, ellis-tang, threshold-delta2,us} 
an explicit delta, the near-threshold behavior is reproduced in lowest orders 
with a monotonically decreasing phase shift. 
The turnaround can at best be achieved if a nominally higher-order
effect provides an opposite contribution to the lower orders.
This is not a problem when this region is considered beyond
the range of EFT, but needs to be addressed as we extend this range,
as done in Refs. \cite{pascalutsa,morepascalutsa1,morepascalutsa2,us}.
The attraction in this channel has long been identified as due to the Roper, 
thanks to its relatively low position and large width.

In this article we incorporate the Roper in ChPT, leading to an expansion in 
$\sim (Q, m_\pi)/\Sigma$, where 
$Q$ now includes $\rho$ as well. 
We continue to refer to this EFT as ChPT because it still relies
on expansions in the quark masses and in momenta.
We illustrate its effect in elastic $\pi N$ scattering. 
We consider $E\sim \delta$ and show that the Roper pole diagram is enhanced, 
significantly improving the description of the $P_{11}$ channel at 
the first non-vanishing order, ${\mathcal O}(Q)$. 
We check explicitly that this description is preserved at next order,
${\mathcal O}(Q^2/\mqcd)$.
We refrain in this first approach from pushing the theory to $E\sim \rho$. 
At such energies a resummation is necessary, just like that in the $P_{33}$ 
channel at $E\sim \delta$. 
However, the proximity to the scale where other effects 
($\sigma$, $N^\star(1520)$, $N^\star(1535)$) accumulate is likely to lead to 
slow convergence. Nevertheless, since the Roper lies not far from the delta
and its width is large, its effects are felt long before $E\sim \rho$. 

Aspects of Roper physics ---the $m_\pi$ dependence of its mass and width--- 
have already been considered in ChPT \cite{roper} with an eye to 
lattice extrapolations. The role of the Roper and other resonances on the 
properties of the baryon decuplet has been discussed in 
$SU(3)$ ChPT \cite{milana}.
An early study of the Roper in $\pi N$ scattering appeared in 
Ref. \cite{threshold-delta1}, although no considerations of 
power counting guided the selection of contributions. 
Note that other approaches exist to incorporate the Roper 
(and other resonances) consistently with chiral symmetry and 
field-definition independence.
They are reminiscent of the original approach  \cite{original,bira-thesis} 
to nuclear interactions using a chiral Lagrangian: a pion-nucleon ``kernel'' 
is first derived in ChPT to a certain order and then unitarized, 
for example using the $N/D$ method \cite{unitarization1,moreIRdeltaless} 
or the Bethe-Salpeter equation \cite{unitarization2}.
(Similar approaches based on meson-exchange models include
those in Ref. \cite{mesonexchange}.)
Power counting is not manifest at the amplitude level, but good results 
for pion-nucleon phase shifts are obtained into the Roper region. 
Needless to say, the Roper has long been been shown to be important 
in phenomenological hadronic models \cite{Kmodelsericson}. 
For a recent review of Roper properties, see Ref. \cite{alvarez}.

The EFT contains all interactions allowed by the symmetries of QCD. 
The chiral Lagrangian with pion ($\iv{\pi}$), nucleon ($N$) and delta 
($\Delta$) fields that is required for $\pi N$ scattering up to 
${\mathcal O}(Q^2/\mqcd)$ in the channel of interest is given in 
Refs. \cite{jenkins, hemmertdelta, bira-thesis, us}. 
We adopt for definiteness the chiral Lagrangian in the form of Ref. \cite{us},
and enlarge it by introducing a heavy-Roper field ($R$) with the 
same quantum numbers as the nucleon. 
Since approximate $SU(2)_L \times SU(2)_R$ chiral symmetry can be accounted 
in EFT through a non-linear realization based on unbroken isospin, 
the technology to construct interactions involving this field is the same 
as for the nucleon. 
The Lagrangian terms can be organized according to the chiral index 
$\nu = d + m + n_{\delta} + n_{\rho} + f/2 - 2 \geqslant 0$ of an interaction, 
where $d$, $m$, $n_{\delta}$, $n_{\rho}$ and $f$ count derivatives, 
powers of $m_\pi$, powers of $\delta$, powers of $\rho$, and 
number of baryon fields, respectively. 
In the following we will need explicitly only the lowest-index Lagrangian,
\beq
\begin{split}
\mathcal{L}^{(0)} &= 2 f_\pi^2 \iv{D}^2
- \frac{m_{\pi}^2 }{2} \frac{\iv{\pi}^2}{\left(1+ \iv{\pi}^2/4f_\pi^2 \right)}
+  N^{\dagger} i \mathscr{D}_0 N 
+ g_A N^{\dagger} \iv{\tau} \vec{\sigma} N \bm{\cdot} \cdot \vec{\iv{D}} \\ 
& \quad {} + \Delta^{\dagger} \left( i \mathscr{D}_0 - \delta \right) \Delta
+ h_A \left( N^{\dagger} \iv{T} \vec{S} \Delta+ H.c. \right) 
      \bm{\cdot} \cdot \vec{\iv{D}} \\ 
& \quad {} +  R^{\dagger} \left( i \mathscr{D}_0 - \rho \right) R
+ g'_A \left(N^{\dagger} \iv{\tau} \vec{\sigma} R+ H.c. \right)  
      \bm{\cdot} \cdot \vec{\iv{D}} +\cdots  \; ,
\label{eqn:lag0}
\end{split}
\eeq
and its first correction
\beq
\begin{split}
\mathcal{L}^{(1)} & =
N^{\dagger} \left(\frac{\vec{\mathscr{D}}^2}{2m_N}
+B_2 \vec{\iv{D}}\cdot \vec{\iv{D}}
+B_3 \varepsilon_{abc}\varepsilon_{ijk}D_{ai}D_{bj}\tau_c \sigma_k \right) N 
+ \frac{1}{2m_N}\Delta^{\dagger} \vec{\mathscr{D}}^2\Delta 
+ \frac{1}{2m_N}R^{\dagger} \vec{\mathscr{D}}^2R
\\ 
& \quad {} - \frac{g_A}{2 m_N} 
 \left( i N^{\dagger}\iv{\tau}\vec{\sigma}\cdot\vec{\mathscr{D}}N + H.c.\right)
 \bm{\cdot}\iv{D}_0 
- \frac{h_A}{m_N} 
 \left( i N^{\dagger} \iv{T} \vec{S}\cdot\vec{\mathscr{D}}\Delta + H.c.\right) 
 \bm{\cdot}\iv{D}_0 \\
& \quad {} - \frac{g_A'}{m_N}
  \left(i N^\dagger\iv{\tau}\vec{\sigma}\cdot\vec{\mathscr{D}}R + H.c.\right)
 \bm{\cdot}\iv{D}_0 + \cdots .
\label{eqn:lag1}
\end{split}
\eeq
Here
$\iv{D}_{\mu}= \partial_{\mu} \iv{\pi}/2 f_\pi +\ldots $ and 
$\mathscr{D}_{\mu}=
\partial_{\mu}+i\iv{t}^{(I)}\cdot(\iv{\pi}\times\iv{D}_{\mu})/f_\pi$ 
are the pion and baryon chiral-covariant derivatives; 
$\iv{t}^{(I)}$ is the isospin generator in a representation of isospin $I$;
$\vec{\sigma}$ and $\iv{\tau}=2\iv{t}^{(1/2)}$ 
are the Pauli matrices in spin and isospin; 
$\vec{S}$ and $\iv{T}$ are $2 \times 4$ transition matrices in spin and 
isospin, normalized so that 
$S_i {S_j}^{\dagger} =\left(2 \delta_{ij} - i\epsilon_{ijk} \sigma_k \right)/3$
and analogously for $\iv{T}$; 
$g_A$, $h_A$, and $g'_A$ are the leading coupling constants of the pion with 
the nucleon, nucleon-delta, and nucleon-Roper, respectively; 
and $B_{2,3}$ are low-energy constants (LECs) of ${\mathcal O}(1/\mqcd)$.
For simplicity we work here in the isospin-symmetric limit,
although (generically small) isospin breaking can be introduced
along the lines of Ref. \cite{bira-thesis}.

Contributions to an arbitrary low-energy process can be ordered in powers of 
$Q/M_{hi}\sim M_{lo}/M_{hi}$, where
$M_{lo}\sim m_\pi \sim \delta \sim \rho <M_{hi}\sim \Sigma \simlt \mqcd$. 
Throughout the low-energy region the standard ChPT power counting 
\cite{weinberg79} applies. In the near-threshold region, the theory is 
purely perturbative in powers of $Q/M_{hi}$. 
In a region $|E - \delta| = \mathcal{O}(\delta^3/M_{hi}^2)$ around
the delta pole, there is a ``kinematic'' fine-tuning: 
one-delta-reducible contributions are enhanced and the delta self-energy 
needs to be resummed \cite{us} 
(and, for a slightly different, earlier version, \cite{pascalutsa}).
At $\mathcal{O}(M_{hi}^2/Q)$ the $\pi N$ amplitude is non-zero only in 
the $P_{33}$ channel, where it has a standard Breit-Wigner form
with a constant width. 
At $\mathcal{O}(Q)$, the width acquires an energy dependence, 
and an energy-dependent background appears in all $S$ and $P$ waves, 
stemming from standard tree diagrams without a delta pole. 
Using $m_\pi=139$ MeV, $m_N=939$ MeV, $f_\pi=92.4$ MeV, and $g_A = 1.29$ 
as input, we find \cite{us} a very good fit 
to $P_{33}$ phase shifts \cite{gwpwa} with $\delta=305$ MeV and $h_A=2.92$. 
Thus, $h_A/g_A$ is not far from the large-$N_c$ value 
$3/\sqrt{2}$ \cite{largenc}. 
The resulting $P_{13}$ and $P_{31}$ phases are equal \cite{us}, 
in qualitative, and at low energies reasonably quantitative, agreement 
with the corresponding observed phase shifts \cite{gwpwa}. 
Thus, ChPT can be successfully extended beyond $E\sim \delta$.

\begin{figure}
\centering
\includegraphics[scale=1]{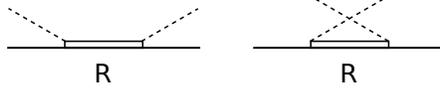}
\caption[Roper trees]{
Roper pole and crossed tree diagrams. 
A double solid line marked ``R'' represents a Roper, 
a single solid line a nucleon, and a dashed line a pion.}
\label{fig:ropertrees}
\end{figure}

Now consider the effects of the Roper. The dominant Roper contributions
will generically be the tree diagrams in \reffig{fig:ropertrees}. 
More complicated diagrams, involving more derivatives and/or loops, 
should be suppressed by powers of $Q/M_{hi}$, and, barring fine-tuning, 
be of higher orders. 
The dominant diagrams themselves, if the Roper is integrated out, would 
contribute to the ``seagull`` counterterms, $\pi \pi \bar{N}N$, 
in the sub-leading Lagrangians of the Roperless EFT. 
Their contributions would appear together with
others of
${\mathcal O}(Q^2/M_{hi})$. 
The Roper crossed diagram is indeed of this order: it decreases with
energy and is, at $E\sim \delta$,
\beq
\frac{Q^2}{E+\rho} \sim \frac{Q^2}{\delta+\rho}\sim \frac{Q^2}{M_{hi}} \; .
\eeq
However, the Roper pole diagram increases in importance as the energy increases
and at $E\sim \delta$ is 
\beq
\frac{Q^2}{E-\rho} \sim \frac{Q^2}{\delta-\rho}\sim \frac{Q^2}{M_{lo}} \;.
\label{PCroperpole}
\eeq
Because the Roper and the delta are not widely separated, the first Roper 
contribution in the delta region is not suppressed by a large scale $M_{hi}$, 
but by the small scale $M_{lo}$. The size of this contribution is 
${\mathcal O}(Q)$, not ${\mathcal O}(Q^2/\mqcd)$. 
It is comparable to the nucleon pole and crossed diagrams,
\beq
\frac{Q^2}{E} \sim \frac{Q^2}{\delta}\sim \frac{Q^2}{M_{lo}} \;,
\eeq
and to the delta crossed diagram
\beq
\frac{Q^2}{E+\delta} \sim \frac{Q^2}{2\delta}\sim \frac{Q^2}{M_{lo}} \;.
\eeq
Near threshold, $E \sim m_\pi$, this counting overestimates
the Roper and delta effects 
due to the numerical difference between $\rho-m_\pi$ and $\delta + m_\pi$,
on the one hand, and $m_\pi$ on the other. 

As $E$ increases further and enters a window of size 
$|E - \rho| = \mathcal{O}(\rho^3/M_{hi}^2)$ around the Roper pole, 
the Roper-pole diagram becomes ${\mathcal O}(M_{hi}^2/Q)$ and
the Roper self-energy is comparable to $|E - \rho|$. 
In this window one-Roper-irreducible diagrams require the same treatment 
as for the delta in Ref. \cite{us}. 
At ${\mathcal O}(M_{hi}^2/Q)$, the Roper self-energy is 
made of one-loop diagrams that should be resummed.
Already at ${\mathcal O}(Q)$, however, complicated two-loop diagrams
appear which account for the Roper decay into two pions.
Although this decay mode is significant, it is still less 
important than decay into a single pion \cite{pdg}. 
This suggests that the EFT expansion could work even in this region,
but we defer a detailed study of $\pi N$ scattering at
the less favorable energy 
$E \sim \rho$ to a later publication.

Outside this window, the Roper pole diagram should capture the dominant part 
of the rise of the resonance, and is the only Roper effect that needs to be 
considered to ${\cal O}(Q)$, or N$^2$LO. 
In $\pi N$ scattering, it contributes only to the 
$P_{11}$ channel. The inclusion of an explicit Roper does not modify the 
amplitudes in the other $P$-wave channels, and therefore does not spoil 
the good description found in Ref. \cite{us}.
This special treatment of the Roper is different from the way the delta is 
dealt with in Ref. \cite{us}. It enables us to improve the description of 
$P_{11}$ over Ref. \cite{us} with minimal complication, provided that one 
stays below the Roper, $E \sim \delta < \rho$.
In this region an extension to higher orders poses no significant
problems, although at some order of course two-loop diagrams also appear.
Whether the relatively slow convergence of the width will
translate into a slow convergence of the full loop expansion
at low energies remains to be seen.

\begin{figure}
\centering
\includegraphics[scale=1]{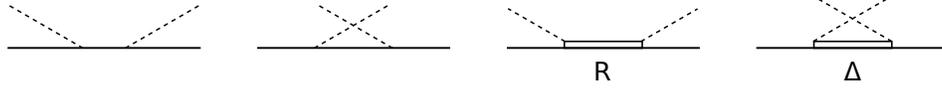}
\caption[NNLO trees]{
N$^2$LO 
diagrams. Vertices have $\nu = 0$.
A double solid line marked ``$\Delta$'' represents a delta isobar;
other notation as in \reffig{fig:ropertrees}.} 
\label{fig:LOtrees}
\end{figure}

To see the dominant Roper effects,
we work with $\pi N$ scattering in the CM frame, 
and denote by $k$ the magnitude of the pion momentum,
by $\omega = \sqrt{k^2 + m_\pi^2}$ its energy,
and by $W_\text{CM}\equiv m_N+E$ the total CM energy.
For $E\sim \delta$,
the first non-vanishing contributions in the $P_{11}$ channel are given 
by the nucleon pole and crossed diagrams,
the delta crossed diagram, and the Roper pole diagram 
(see \reffig{fig:LOtrees}), all constructed entirely from $\mathcal{L}^{(0)}$, 
Eq. (\ref{eqn:lag0}). 
We find for the first non-vanishing contributions 
to the $T$ matrix in the $P_{11}$ channel
\beq
T_{P_{11}}^\text{N$^2$LO}=
-\frac{g_A^2}{3 \pi f_\pi^2} k^3\mathcal{N}(k)
\left[\frac{1}{E}- \left(\frac{\sqrt{2}h_A}{3g_A}\right)^2 \frac{1}{E+\delta}  
+ \frac{9}{8} \left(\frac{g'_A}{g_A} \right)^2\frac{1}{E-\rho} 
\right] \, ,
\label{eqn:Tp131} 
\eeq
where 
\beq
\mathcal{N}(k) \equiv 
\frac{1+ \left(E-\omega\right)/m_N}{1+ E/m_N} 
\label{eqn:defN}
\eeq
is a kinematic coefficient. 
Here the nucleon and delta terms are the same as in Ref. \cite{us}. 
Since the relative coefficient of the delta contribution is $\simeq 1$, 
the nucleon contribution is numerically larger
and leads (in the absence of the Roper term) to repulsive phase shifts 
throughout the low-energy region. 
As long as $(g'_A/g_A)^2$ is not too large, 
this situation survives near threshold.
As $E$ increases, however, while (apart from the overall $k^3$) the nucleon 
and delta contributions decrease in magnitude, 
the Roper contribution increases. 
Since it has sign opposite to the nucleon term as long as $E<\rho$, 
it will eventually overcome the others. 
As long as $(g'_A/g_A)^2$ is not too small, 
this happens at energies $E\sim M_{lo}$.  

\begin{figure}
\centering
\includegraphics[scale=1]{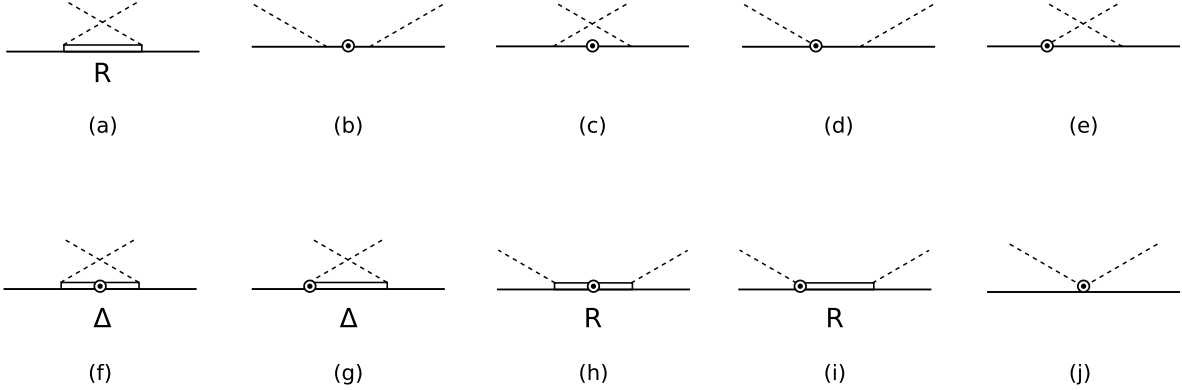}
\caption[N3LO trees]{
N$^3$LO 
diagrams. Once-circled vertices have $\nu = 1$;
other notation as in \reffig{fig:ropertrees} and \reffig{fig:LOtrees}. 
Diagrams with the same topology but different permutation of vertices 
are drawn only once.} 
\label{fig:NLOtrees}
\end{figure}

In next order, ${\cal O}(Q^2/M_{hi})$ or N$^3$LO, 
there are more tree-level contributions, see \reffig{fig:NLOtrees}. 
In addition to the Roper crossed diagram (a), these contributions stem from 
one insertion of terms in $\mathcal{L}^{(1)}$, Eq. (\ref{eqn:lag1}): 
nucleon recoil (b, c) and
the Galilean pion-nucleon vertex correction (d, e) 
in nucleon pole and crossed diagrams;
delta recoil (f) and the Galilean pion-nucleon-delta vertex correction (g) 
in delta crossed diagrams; 
Roper recoil (h) and the Galilean pion-nucleon-Roper vertex correction (i) 
in Roper pole diagrams; 
and pion-nucleon seagulls (j). 
Of these diagrams, (b), (h) and (i) vanish in the CM frame, 
and (c) and (f) do not contribute to the $P_{11}$ partial wave. 
We find
\beq
\begin{split}
T_{P_{11}}^\text{N$^3$LO} &= -\frac{g_A^2}{3 \pi f_\pi^2} k^3\mathcal{N}(k)
\left[ \frac{\omega}{E\, m_N} -
\left(\frac{\sqrt{2}h_A}{3g_A}\right)^2 \frac{2\omega}{(E+\delta)m_N}
-\frac{1}{8} \left(\frac{g'_A}{g_A}\right)^2  \frac{1}{E+\rho}
+ \frac{B}{g_A^2}\right]
\; ,
\label{eqn:Tp131next}
\end{split}
\eeq
where 
\begin{equation}
B= B_3-\frac{B_2}{4} - \frac{1}{4m_N}
\end{equation} 
in terms of the LECs $B_{2,3}$.
The nucleon contributions in Eqs.~\eqref{eqn:Tp131} and \eqref{eqn:Tp131next} 
are in agreement with Ref.~\cite{fettes-deltaless-is}. 
While Eq.~\eqref{eqn:Tp131} agrees with the $\mathcal{O}(Q)$ delta 
contributions in Ref.~\cite{threshold-delta2}, a comparison 
at $\mathcal{O}(Q^2)$ is obscured by the employment in 
Ref.~\cite{threshold-delta2} of off-shell parameters 
and a $\nu=1$ $\pi N \Delta$ coupling, $b_3 + b_8$,
which is removed in our calculation using 
integration by parts and baryonic equations of motion \cite{lensky}.
To this order, our EFT resembles an isobar model
such as that in Ref.~\cite{Kmodelsericson},
the main difference lying on the $B$ term.
Such term
accounts for the short-range physics 
not considered explicitly in ChPT. 
Although unknown, it is not completely free:
for the EFT expansion to be sensible, $B$ is 
expected to be naturally sized, 
$B \sim 1/M_{hi}$.
Pion loops and further LECs appear in sub-leading orders.

Expanding also the phase shifts in powers of $Q/M_{hi}$, they can be 
extracted from Eqs. \eqref{eqn:Tp131} and \eqref{eqn:Tp131next} 
in such a way that unitarity is preserved perturbatively:
$\theta_{P_{11}}^{\text{N$^2$LO}} = T_{P_{11}}^\text{N$^2$LO}/2$ 
and $\theta_{P_{11}}^{\text{N$^3$LO}} = T_{P_{11}}^\text{N$^3$LO}/2$. 
If we use the same nucleon and delta parameters determined at N$^2$LO
in Ref. \cite{us},
we have just two undetermined parameters at N$^2$LO, $\rho$ and $g'_A$, 
and one more at N$^3$LO, $B$. 
We fit our results to the $P_{11}$ phases from the 
energy-dependent solution of the phase-shift analysis (PSA) 
by the George Washington (GW) group~\cite{gwpwa}.

The $P_{11}$ phase shifts have an interesting feature:
they almost vanish until reaching the sign-flip point, $E \sim \delta$, 
beyond which they become attractive. The phase is only about $-1^\circ$ 
at the lowest point of the ``dip''. This suggests that, for a wide kinematic 
window, the repulsion contributed by the nucleon nearly cancels the 
attraction by the delta and Roper. This sort of cancellation stands out
among the low-energy $S$ and $P$ partial waves.
While the sign-flip point constrains $\rho$ and $g_A'$ at N$^2$LO 
(Eq.~\eqref{eqn:Tp131}), the near cancellation for a wide window implies 
that the constraint is not restrictive at all, within an error of $1^\circ$. 
Thus, we expect a significant uncertainty in determining $\rho$ or $g_A'$.
We reduce this uncertainty by fitting points 
on the right side of the sign-flip point,
from $W_\text{CM} = 1250$ MeV to $1300$ MeV,
where the phase shifts become appreciable.
Since the energy-dependent solution of the GW PSA does not have error bars, 
each input PSA point is fitted with the same weight, and only the central 
values of the LECs are shown below.

\begin{figure}
\centering
\includegraphics[scale=1]{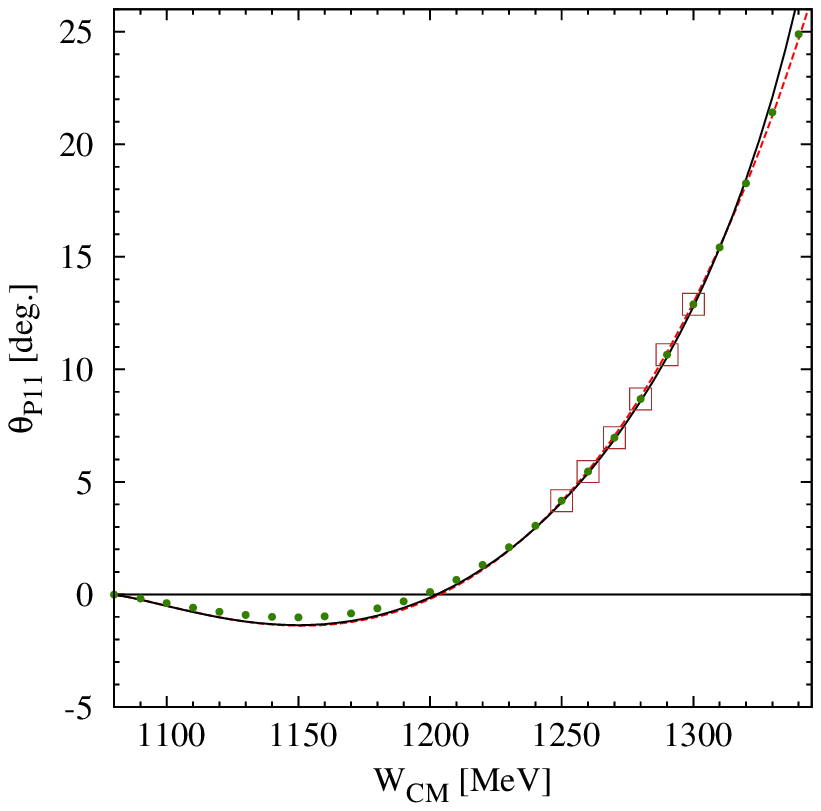}
\includegraphics[scale=1]{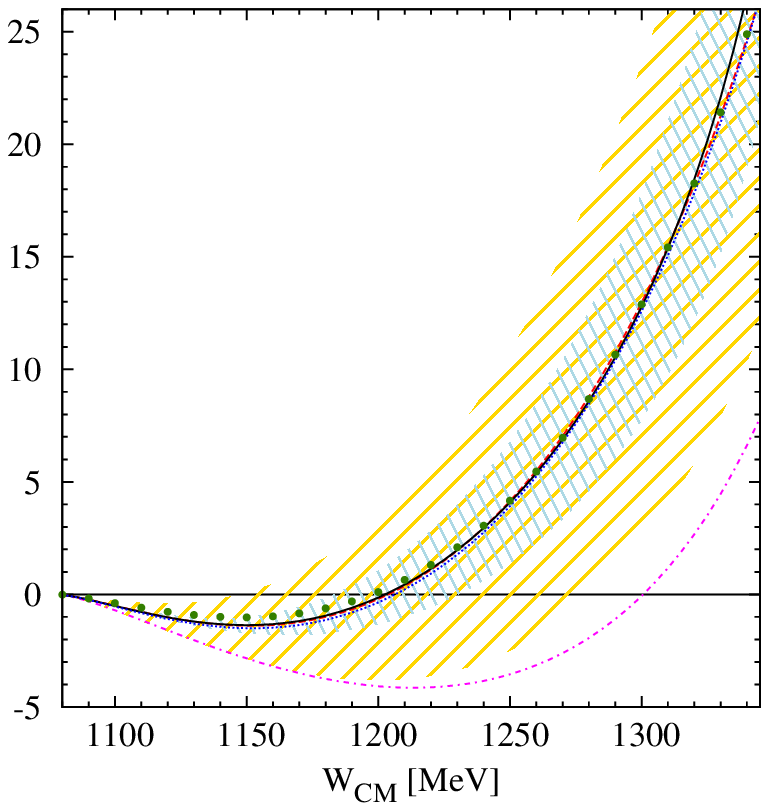}
\caption{\label{fig:p11}
The $P_{11}$ phase shift $\theta_{P_{11}}$ as a function of $W_\text{CM}$, 
the CM energy including the nucleon mass. 
LO and NLO vanish in this channel; 
the N$^2$LO EFT fit 
is given by the (red) dashed line and the N$^3$LO,
by the (black) solid line.
The green dots are the results of the GW phase-shift analysis \cite{gwpwa}. 
On the left plot, the PSA points used as input in the EFT fit
are marked by square boxes.
On the right plot, the theoretical error band at N$^2$LO is indicated
by a (gold) forward hatched area, and at N$^3$LO by a (light blue) backward
hatched area.
We also show as, respectively, (magenta) dot-dashed and (blue) dotted curves
the N$^2$LO 
and N$^3$LO 
results using as input the Breit-Wigner mass and width
from the GW analysis \cite{gwpwa}.
}
\end{figure}

The results for the $P_{11}$ phase shifts are shown in the left plot of 
\reffig{fig:p11}. 
For most of the plot, the N$^3$LO curve coincides with the N$^2$LO 
curve. It fits slightly worse on the uphill to the Roper resonance,
but does slightly better
in the region we take the PSA inputs.
The two curves agree with each other and data much more than one would
expect. 
The theoretical error of the EFT amplitudes
can be estimated as follows: 
(\textit{i}) at N$^2$LO,
\beq
\Delta T_{P_{11}}^\text{N$^2$LO} = 
\pm
\frac{k^2 \mathcal{N}(k)}{3\pi f_\pi^2}
\frac{k}{M_{hi}} \, ,
\label{error2}
\eeq
which can be interpreted as the $\pi N$ seagull terms with a ``natural'' size;
(\textit{ii}) at N$^3$LO,
\beq
\Delta T_{P_{11}}^\text{N$^3$LO} = 
\pm \frac{k^2 \mathcal{N}(k)}{3\pi f_\pi^2} \frac{k^2}{M_{hi}^2} \, .
\label{error3}
\eeq
The theoretical errors, with $M_{hi} = 600$ MeV,
are indicated by shaded areas in the right plot of FIG.~\ref{fig:p11}.

From the fit we extract values for $\rho$, $g'_A$, and $B$. 
These are given in the first two rows of
TABLE \ref{tbl:lecs}, together with values 
for $\delta$ and $h_A$ found in Ref. \cite{us}, 
and the value of $g_A$ used as input. 
These are the values that give 
the curves in the left panel of \reffig{fig:p11}.
In order to have an estimate of the uncertainty in a LEC,
we consider its variation so that 
the EFT curves roughly stay, up to 1300 MeV, within the theoretical error band 
in the right panel of \reffig{fig:p11},
while other LECs are fixed at central values.
At N$^3$LO, we find 
$g'_A = 0.32^{+0.14}_{-0.18}$,
$\rho = 470^{+250}_{-40}$ MeV,
and $B= -2.3^{+0.7}_{-0.7}$ GeV$^{-1}$.
Of course, the errors in the parameters are correlated
and these variations are just an illustration of the 
uncertainty in the parameter-space region that generates the 
extreme curves. 
These results are compared to values, in the last row,
from the assumption that the nucleon, delta, and Roper form a reducible 
representation of chiral symmetry with maximal mixing \cite{roperQCD}.
At N$^3$LO there is agreement better than $15\%$ for central values,
well within the errors of the N$^3$LO extraction.
(Note that the signs of $h_A$ and $g'_A$ cannot be extracted from
our analysis, and were simply chosen to be the same
as in Ref. \cite{roperQCD}.)

\begin{table}
\caption{\label{tbl:lecs}
Low-energy constants
appearing in the $P_{11}$ channel up to N$^3$LO encoding
nucleon ($g_A$), delta ($h_A$ and $\delta$, in MeV),
Roper ($g_A'$ and $\rho$, in MeV), and higher-energy
($B$, in GeV$^{-1}$) properties.
The values 
for $g_A'$, $\rho$ and $B$
extracted from the EFT fits at N$^2$LO and N$^3$LO 
(where $g_A$, $h_A$, and $\delta$
were used as input)
are labeled ``EFT''.
In the rows labeled ``GW'', $\rho$ and $g_A'$ are extracted 
from the Breit-Wigner parameters for the Roper \cite{gwpwa}.
The values when the nucleon, delta and Roper
form a reducible representation of chiral symmetry 
with maximal mixing \cite{roperQCD} are labeled ``Chiral rep''.
}
\begin{tabular}{|c||c|c|c|c|c|c|}
\hline
& $g_A$ & $h_A$ & $\delta$ (MeV) & $g'_A$ & $\rho$ (MeV) & $B$ (GeV$^{-1}$) \\
\hline
\hline
N$^2$LO EFT       & 1.29 & 2.92 & 305 & 1.06 & 690 & ---  \\
\hline
N$^3$LO EFT        & 1.29 & 2.92 & 305 & 0.32 & 470 & $-2.3$ \\
\hline
N$^2$LO GW       & 1.29 & 2.92 & 305 & 0.54 & 550 & ---   \\
\hline
N$^3$LO GW       & 1.29 & 2.92 & 305 & 0.54 & 550 & $-1.6$ \\
\hline
Chiral rep & 1.33 & 2.82 & 292 & 0.33 & 527 & --- \\
\hline
\end{tabular}
\end{table}

Although the two fitted phase-shift curves almost coincide, 
we should not conclude that the EFT expansion in $P_{11}$ converges 
very rapidly. 
The relatively large variation of $\rho$ and $g_A'$ from N$^2$LO to 
N$^3$LO shows the significant impact of the N$^3$LO correction.
Because $B$ is still naturally sized, $\mathcal{O}(1/\mqcd)$, 
the N$^3$LO amplitude (Eq.~\eqref{eqn:Tp131next}) is one order 
of magnitude smaller than the magnitude of either the repulsive or 
the attractive term at N$^2$LO (Eq.~\eqref{eqn:Tp131}). 
But, in order for the sum of two orders to nearly vanish below 
$E \simeq \delta$,
the N$^2$LO and its subleading corrections need to be numerically comparable,
which indicates slow convergence of the EFT expansion in this channel. 
However, the overall convergence is
reasonable in the sense that
the N$^3$LO $P_{11}$ is still
one order smaller than, e.g., the $P_{33}$ N$^2$LO amplitude \cite{us}.

The large change 
in Roper parameters from one order to the next
is consequence of 
fitting them away from
the Roper resonance region, which is most sensitive to them.
In order to have another estimate of uncertainties and convergence, we
consider an alternative way to determine the Roper parameters.
The leading $\pi N$ partial width of the Roper is
\beq
\Gamma_{\pi N}^{(0)}(E)=\frac{3}{8 \pi} \left(\frac{g_A'}{f_\pi}\right)^2
\mathcal{N}(k_\rho) k_\rho^3 \, ,
\eeq
where $k_\rho$ is the CM momentum when the Roper is on-shell. 
When we fit $\rho$ and $g_A'$ to the Breit-Wigner mass and width
from GW PSA \cite{gwpwa},
and at N$^3$LO $B$ from the same phases as before,
we obtain two extra curves shown in the right panel of \reffig{fig:p11}.
The parameters obtained this way
are shown in the rows of TABLE \ref{tbl:lecs} labeled ``GW''. 
Although the values of $\rho$ and $g_A'$ 
extracted from the Breit-Wigner resonance parameters
appear quite different from those fitted at N$^2$LO,
the phenomenological curve
is in the vicinity of the N$^2$LO error band,
which confirms the difficulty in
extracting $\rho$ and $g_A'$ from the EFT amplitude at N$^2$LO. 
The smaller error band at N$^3$LO suggests a more reliable extraction of LECs: 
$\rho$ indeed has come closer to the Breit-Wigner mass 
or the real part of the Roper pole position. 
The large variation in the ``GW'' curves from  N$^2$LO to N$^3$LO
is a reflection of the importance of the background (here represented 
by the LEC $B$), which in turn is another consequence of slow convergence. 
Note that we use this procedure based on Breit-Wigner parameters
only as an illustration that our Roper N$^3$LO values might
survive an extension of the theory to the Roper region;
we cannot, and do not, make claims about the success of an eventual fit
in this region.


Although we do not aim at a precise description of the threshold region,
we extract for completeness the corresponding scattering volume at N$^3$LO:
\beq
\begin{split}
a_{P_{11}} &=
-\frac{g_A^2}{6 \pi f_\pi^2 m_\pi}
\left\{1
- \left(\frac{\sqrt{2}h_A}{3g_A}\right)^2
          \frac{m_\pi/\delta}{1+m_\pi/\delta}
\frac{1 + 2m_\pi/m_N}{1+m_\pi/m_N}
\right. \\
&\left. \qquad\qquad\qquad
-\left[\frac{5}{4} \left(\frac{g'_A}{g_A}\right)^2\frac{m_\pi}{\rho}
        \frac{1+4m_\pi/5\rho}{1-m_\pi^2/\rho^2}
       -\frac{m_\pi B}{g_A^2}
       \right]\left(1+\frac{m_\pi}{m_N}\right)^{-1}
\right\} \, .
\end{split}
\label{eqn:apnd} 
\eeq
The delta and Roper contributions are suppressed by factors of 
$m_\pi/\delta$ and $m_\pi/\rho$, respectively. 
With the N$^3$LO EFT values of $\rho$ and $g'_A$ in TABLE \ref{tbl:lecs}, 
the delta, Roper and seagull counterterm contributions are, respectively, 
$-40\%$, $-2.7\%$ and $-17\%$ of the nucleon's, and 
$a_{P_{11}}=
-0.081 \, m_\pi^{-3}$, in a good agreement 
with $a_{P_{11}}=-(0.0799\pm 0.0016)m_\pi^{-3}$ 
found in a PSA based on low-energy data \cite{matsinos}.
Of course, the successful description near threshold owes more to the seagull 
than to the Roper.
The scattering length can be fitted in Roperless, deltaless ChPT
with an effective $B_{eff}$ that accounts for the delta, Roper,
and higher-energy contributions. 
It is reassuring that parameters fitted at $E\sim \delta$
produce nearly the same value for $B_{eff}$,
the expression of 
which to N$^3$LO in our EFT can be directly read off Eq. \eqref{eqn:apnd}.

In summary, we have argued that the Roper can profitably be included
in ChPT.
As a first study, 
we focused on $\pi N$ scattering 
below the Roper resonance,
in the region around the delta.
We proposed the promotion of the Roper pole diagram over its crossed 
counterpart. Using this counting scheme, we found a good description 
of the $P_{11}$ phase shifts throughout the low-energy region
already at lowest non-vanishing order. 
The smallness of the $P_{11}$ phase shifts at low energies
makes difficult a reliable extraction of LECs from the EFT amplitude at 
this order,
but at next order the best fit gives 
LECs close to those predicted by a reducible chiral representation
with maximal mixing.
The cancellations necessary for small $P_{11}$ phase shifts
suggest a slow convergence at low orders 
in the $P_{11}$ channel, 
but it does not spoil overall convergence, when all
partial waves, including those of more natural size,
are taken into account.

\acknowledgments
We thank Silas Beane for useful comments.
We are grateful to the following institutions for hospitality while this work 
was being carried out: the Kernfysisch Versneller Instituut 
at Rijksuniversiteit Groningen (UvK), 
the National Institute for Nuclear Theory at the University of Washington 
(BwL, UvK), and the University of Arizona (BwL). 
This work was supported by the US DOE 
under contracts 
DE-AC05-06OR23177 (BwL) and 
DE-FG02-04ER41338 (UvK). 
This work is coauthored by Jefferson Science Associates, LLC 
under US DOE Contract No. DE-AC05-06OR23177.


\begin{thebibliography}{99}

\bibitem{weinberg79} 
S.~Weinberg, 
\textit{Physica} \textbf{96A} (1979) 327;
J.~Gasser and H. Leutwyler,
\textit{Ann. Phys.} \textbf{158} (1984) 142;
\textit{Nucl. Phys.} \textbf{B250} (1985) 465.

\bibitem{BerKaiMei}
V.~Bernard, N.~Kaiser, and U.-G.~Mei{\ss}ner,
\textit{Int. J. Mod. Phys.} \textbf{E4} (1995) 193;
V.~Bernard,
\textit{Prog. Part. Nucl. Phys.} \textbf{60} (2008) 82.

\bibitem{Caprini:2005zr}
I.~Caprini, G.~Colangelo, and H.~Leutwyler,
\textit{Phys.\ Rev.\ Lett.} {\bf 96} (2006) 132001.

\bibitem{gasser}
J. Gasser, M.E. Sainio, and A. \v{S}varc,
\textit{Nucl. Phys.} \textbf{B307} (1989) 779;
E. Jenkins and A.V. Manohar,
\textit{Phys. Lett.} \textbf{B255} (1991) 558.

\bibitem{gwpwa}
R.A.~Arndt, W.J.~Briscoe, I.I.~Strakovsky, and R.L.~Workman,
\textit{Phys. Rev.} \textbf{C74} (2006) 045205;
R.A.~Arndt \textit{et al.},
The SAID program, \texttt{http://gwdac.phys.gwu.edu/}.

\bibitem{jenkins}  
E.~Jenkins and A.V.~Manohar, 
\textit{Phys. Lett.} \textbf{B259} (1991) 353;
in {\it Effective Field Theories of the Standard Model},
U.-G. Mei{\ss}ner (editor), World Scientific, Singapore (1992);
E. Jenkins,
\textit{Nucl. Phys.} \textbf{B375} (1992) 561.

\bibitem{hemmertdelta} 
T.R.~Hemmert, B.R.~Holstein, and J.~Kambor,
\textit{Phys. Lett.} \textbf{B395} (1997) 89;
\textit{J. Phys.} \textbf{G24} (1998) 1831.

\bibitem{bira-thesis}
U.~van Kolck, 
Ph.D dissertation, U. of Texas (1993);
C.~Ord\'o\~nez, L.~Ray, and U.~van Kolck, 
\textit{Phys. Rev. Lett.} \textbf{72} (1994) 1982;
\textit{Phys. Rev.} \textbf{C53} (1996) 2086;
U.~van Kolck, 
\textit{Phys. Rev.} \textbf{C49} (1994) 2932;
T.D.~Cohen, J.L.~Friar, G.A.~Miller, and U.~van Kolck,
\textit{Phys. Rev.} \textbf{C53} (1996) 2661.

\bibitem{pascalutsa} 
V.~Pascalutsa and D.R.~Phillips, 
\textit{Phys. Rev.} \textbf{C67} (2003) 055202;
V.~Pascalutsa,
\textit{Prog. Part. Nucl. Phys.} \textbf{61} (2008) 27.

\bibitem{us}
Bingwei Long and U. van Kolck,
\textit{Nucl. Phys.} \textbf{A840} (2010) 39.

\bibitem{morepascalutsa1} 
V.~Pascalutsa and M.~Vanderhaeghen, 
\textit{Phys. Rev. Lett.} \textbf{94} (2005) 102003;
\textit{Phys. Rev. Lett.} \textbf{95} (2005) 232001;
\textit{Phys. Rev.} \textbf{D73} (2006) 034003;
\textit{Phys. Rev.} \textbf{D77} (2008) 014027.

\bibitem{morepascalutsa2} 
V.~Pascalutsa, M.~Vanderhaeghen, and S.N.~Yang,
\textit{Phys. Rept.} \textbf{437} (2007) 125.

\bibitem{origroper}
L.D. Roper,
\textit{Phys. Rev. Lett.} \textbf{12} (1964) 340.

\bibitem{algreal}
S. Weinberg,
\textit{Phys. Rev.} \textbf{177} (1969) 2604.

\bibitem{roperQCD}
S.R. Beane and U. van Kolck,
\textit{J. Phys.} \textbf{G31} (2005) 921.

\bibitem{mojzis}
M.~Moj\u{z}i\u{s},
\textit{Eur. Phys. J.} \textbf{C2} (1998) 181.

\bibitem{fettes-deltaless-is}
N.~Fettes, U.-G.~Mei{\ss}ner, and S.~Steininger,
\textit{Nucl. Phys.} \textbf{A640} (1998) 199;
N.~Fettes and U.-G.~Mei{\ss}ner,
\textit{Nucl. Phys.} \textbf{A676} (2000) 311;
\textit{Nucl. Phys.} \textbf{A693} (2001) 693.

\bibitem{otherdeltaless}
T.~Becher and H.~Leutwyler,
\textit{JHEP} \textbf{0106} (2001) 017.

\bibitem{moreIRdeltaless}
J.M. Alarc\'on, J. Mart\'in Camalich, J.A. Oller, and L. Alvarez-Ruso,
\texttt{arXiv:1102.1537}.

\bibitem{threshold-delta1}
A.~Datta and S.~Pakvasa,
\textit{Phys. Rev.} \textbf{D56} (1997) 4322.

\bibitem{ellis-tang}
P.J.~Ellis and H.-B.~Tang,
\textit{Phys. Rev.} \textbf{C57} (1998) 3356;
K.~Torikoshi and P.J.~Ellis,
\textit{Phys. Rev.} \textbf{C67} (2003) 015208.

\bibitem{threshold-delta2}
N.~Fettes and U.-G.~Mei{\ss}ner, 
\textit{Nucl. Phys.} \textbf{A679} (2001) 629.

\bibitem{roper}
B. Borasoy, P.C. Bruns, U.-G. Mei{\ss}ner, and R. Lewis, 
\textit{Phys. Lett.} \textbf{B641} (2006) 294;
D. Djukanovic, J. Gegelia, and S. Scherer,
\textit{Phys. Lett.} \textbf{B690} (2010) 123.

\bibitem{milana}
M.K. Banerjee and J. Milana,
\textit{Phys. Rev.} \textbf{D54} (1996) 5804.

\bibitem{original}
S. Weinberg,
\textit{Phys. Lett.} \textbf{B251} (1990) 288;
\textit{Nucl. Phys.} \textbf{B363} (1991) 3.

\bibitem{unitarization1}
U.-G. Mei{\ss}ner and J.A. Oller,
\textit{Nucl. Phys.} \textbf{A673} (2000) 311.

\bibitem{unitarization2}
M.F.M. Lutz and E.E. Kolomeitsev,
\textit{Nucl. Phys.} \textbf{A700} (2002) 193.

\bibitem{mesonexchange}
M. D\"oring, C. Hanhart, F. Huang, S. Krewald, and U.-G. Mei{\ss}ner,
\textit{Nucl. Phys.} \textbf{A829} (2009) 170;
H. Kamano, S.X. Nakamura, T.-S.H. Lee, and T. Sato,
\textit{Phys. Rev.} \textbf{C81} (2010) 065207.

\bibitem{Kmodelsericson}
E.~Oset, H.~Toki, and W.~Weise,
\textit{Phys. Rept.} \textbf{83} (1982) 282;
T.~Ericson and W.~Weise,
\textit{Pions and Nuclei}, Clarendon Press, Oxford, (1988).

\bibitem{alvarez}
L. Alvarez-Ruso,
arXiv:1011.0609 [nucl-th].

\bibitem{largenc}
A. Manohar, 
hep-ph/9802419.

\bibitem{pdg}
K.~Nakamura  [Particle Data Group],
\textit{J. Phys.} {\bf G37} (2010) 075021.

\bibitem{lensky}
Bingwei Long and V. Lensky, 
\textit{Phys. Rev.} \textbf{C83} (2011) 045206.

\bibitem{matsinos}
E. Matsinos, W.S. Woolcock, G.C. Oades, G. Rasche, and A. Gashi,
\textit{Nucl. Phys.} \textbf{A778} (2006)  95.

\end{thebibliography}
\end{document}